# *A new family of solvable Pearson-Dirichlet random walks*

Gérard Le Caër

*Institut de Physique de Rennes, UMR UR1-CNRS 6251, Université de Rennes I, Campus de Beaulieu, Bâtiment 11A, F-35042 Rennes Cedex, France*

*E-mail address*: gerard.le-caer@univ-rennes1.fr

*Abstract*

A $n$-step Pearson-Gamma random walk in $\mathbb{R}^d$ starts at the origin and consists of $n$ independent steps with gamma distributed lengths and uniform orientations. The gamma distribution of each step length has a shape parameter $q > 0$. Constrained random walks of $n$ steps in $\mathbb{R}^d$ are obtained from the latter walks by imposing that the sum of the step lengths is equal to a fixed value. Simple closed-form expressions were obtained in particular for the distribution of the endpoint of such constrained walks for any $d \geq d_0$ and any $n \geq 2$ when $q$ is either $q = \frac{d}{2} - 1$ ($d_0 = 3$) or $q = d - 1$ ($d_0 = 2$) (J. Stat. Phys., 140: 728, 2010). When the total walk length is chosen, without loss of generality, to be equal to 1, then the constrained step lengths have a Dirichlet distribution whose parameters are all equal to $q$ and the associated walk is thus named a Pearson-Dirichlet random walk. The density of the endpoint position of a $n$-step planar walk of this type ($n \geq 2$), with $q = d = 2$, was shown recently to be a weighted mixture of $1 + floor(n/2)$ endpoint densities of planar Pearson-Dirichlet walks with $q = 1$ (Stochastics, 82: 201, 2010). The previous result is generalized to any walk space dimension and any number of steps $n \geq 2$ when the parameter of the Pearson-Dirichlet random walk is $q = d > 1$. We rely on the connection between an unconstrained random walk and a constrained one, which have both the same $n$ and the same $q = d$, to obtain a closed-form expression of the endpoint density. The latter is a weighted mixture of $1 + floor(n/2)$ densities with simple forms, equivalently expressed as a product of a power and a Gauss hypergeometric function. The weights are products of factors which depend both on $d$ and $n$ and Bessel numbers independent of $d$.

# *1. Introduction*

To model the rate of infiltration of a species into a possible habitat, Pearson defined a simple planar "random walk" [1-2]. The Pearson random walk starts at the origin and consists of a sequence of $n$ steps with identical lengths taken into uniformly random directions. Pearson sought for the help of the readers of Nature to obtain the probability that after these $n$ steps the walker is "a distance between $r$ and $r+dr$ from his origin O.". An asymptotic solution was rapidly given by Rayleigh [3]. In his reply, Pearson emphasized the mathematical interest of finding exact solutions for $n$ comparatively small. Exact expressions were soon obtained for $n=2$ and for $n=3$. After having been left aside for almost one century, the latter question was reconsidered very recently by mathematicians because of its interest in special function theory and in computer algebra [4-5]. Closed-form expressions were obtained for all moments of the distance travelled from the origin after unit steps for 2-step and 3-step walks [4]. The radial density was reformulated for $n=3$ and derived for $n=4$ in term of hypergeometric functions [5].

The Pearson random walk and its variants find numerous applications in diverse fields such as physics, biology, ecology ([6-8] and references therein). A large family of variants includes correlated random walks, which involve a correlation between successive step orientations. They are used to model the movement of animals, micro-organisms and cells, the dispersal of animals, etc..

In this paper, we focus on two variants of the Pearson random walk with steps of uniform random orientation and unequal sizes which take place in Euclidean spaces $\mathbb{R}^d$ and on random flights performed by a particle in $\mathbb{R}^d$ which are equivalent to the latter walks. In both variants, the step lengths of the walks and the displacements between two changes of orientations of the flights are identically distributed. The common step length distribution is either a gamma distribution with a shape parameter $q$ (appendix A) or a Dirichlet distribution, which is directly defined from the previous gamma law, with all its parameters equal to $q$ (appendix A). These variants are named respectively Pearson-Gamma random walk and Pearson-Dirichlet random walk. Such walks or flights were previously considered in the literature, most often for $q=1$ (exponential distribution) [9-17] but also for $q=2$ [10,18] and for $q=\frac{d}{2}-1$ and $q=d-1$ [19].

Interestingly, a closed-form expression of the probability density function of the endpoint of planar Pearson-Dirichlet walks of $n$ $(\geq 2)$ steps was recently obtained by Beghin and Orsingher for $q = d = 2$ [18] as a weighted mixture of $1 + floor(n/2)$ densities of planar walks with $q = 1$. The aim of the present work is to show that this result can actually be generalized to Pearson-Dirichlet walks of $n$ $(\geq 2)$ steps in $\mathbb{R}^d$, with $q = d$, for any space dimension $d > 1$. First, we shall describe all notations. Second, we shall summarize the exact results obtained for the densities of the endpoint positions of the Pearson-Gamma and Pearson-Dirichlet walks mentioned above, and more particularly the exact results of Beghin and Orsingher [18]. Third, we shall establish general relations which connect a Pearson-Gamma walk and a Pearson-Dirichlet walk which have the same parameters $d$, $n$ and $q$. Fourth, we shall present our initial guess of the endpoint density of a Pearson-Dirichlet walk with $q = d$ and the method we used to derive it completely. Last, we shall show how the previously established relations between Pearson-Gamma and Pearson-Dirichlet walks allow us to prove that the latter density is indeed the exact one.

## *2. Notations*

The Pochhammer symbol, $(a)_r = \Gamma(a+r)/\Gamma(a)$, $(a)_0 = 1$, will be used repeatedly throughout the text. It reduces to an ascending factorial, $(a)_r = a(a+1)..(a+r-1)$, when $r$ is an integer. The largest integer not greater than $x$ is $\lfloor x \rfloor = floor(x)$.

### *2.1 Random variables, Gamma and Dirichlet distributions*

As usual upper-case letters will be used to denote random variables and lower-case letters for the values they take. The mean of a function $f(X)$ of a continuous random variable or of a random vector $X$, whose probability density function (pdf) $p(x)$ is defined in some domain $D$, will be denoted hereafter as $\langle f(X) \rangle = \int_D f(x) p_X(x) dx$. The latter integral is possibly multidimensional. The pdf $p(Y|X = x)$ is the conditional density of a continuous random variable $Y$ given the value $x$ of $X$.

The notation $A \triangleq B$ means that the random variables $A$ and $B$ are identically distributed.

Appendix A gives the pdf and the moments of a gamma distributed random variable, $\gamma(q,\theta)$, with a shape parameter $q$ and a scale parameter $\theta$, and those of a Dirichlet distributed $m$-dimensional random vector, $D_m(\boldsymbol{\alpha}^{(n)})$, with $n = m+1$ and a vector of parameters $\boldsymbol{\alpha}^{(n)} = (\alpha_1,...,\alpha_m,\alpha_n)$. The Dirichlet distributions we will more particularly consider have all their parameters equal to $q$. Finally, $\boldsymbol{U}^{(k)}$ stands for a randomly oriented unit vector in $\mathbb{R}^k$.

## 2.2 Characteristic functions

The characteristic function (c.f.) of a $d$-dimensional random vector $\boldsymbol{R}_A$ is the function $\Phi^A(\boldsymbol{\rho}) = \langle e^{i\boldsymbol{\rho}.\boldsymbol{R}_A} \rangle$ of the $d$-dimensional vector $\boldsymbol{\rho}$. For the spherically symmetric distributions considered here, the c.f. depends only on the modulus $\rho = \|\boldsymbol{\rho}\|$. A specific notation, $\Omega_k(\rho)$, will be used for the characteristic function of $\boldsymbol{U}^{(k)}$: $\Omega_k(\rho) = \langle e^{i\boldsymbol{\rho}.\boldsymbol{U}^{(k)}} \rangle$ (appendix B).

## 2.3 Random walks with Gamma and Dirichlet distributed step sizes

The following notations are used throughout the rest of the text:
1) $d$ is the dimension of the Euclidean space in which the walk or the flight takes place.
2) $n$ is the number of steps.
3) The parameter $\beta = \dfrac{(n-1)(d-1)}{2}$ will simplify the writing of many equations.
4) The variants of the Pearson random walks we investigate are named after their step length distribution:
- a Pearson-Gamma random walk, $PG(d,n,q)$, has $n$ independently and identically distributed (i.i.d.) step lengths, each with a $\gamma(q,1)$ distribution (appendix A). The step lengths are denoted as $S_k$ $(k=1,..,n)$ (figure 1).

- a Pearson-Dirichlet random walk, $PD(d,n,q)$, has $n$ step lengths whose distribution is a Dirichlet distribution $D_m(\mathbf{q}^{(n)})$ with $\mathbf{q}^{(n)} = (q,q,..,q)$ (appendix A). These step lengths, whose sum is constrained to be equal to 1, are denoted as $L_k$ ($k=1,..,n$) (figure 1). A Pearson-Dirichlet random walk may alternatively be considered as a constrained Pearson-Gamma random walk whose shape factor is $q$.

As we are chiefly interested in walks and flights with $q=d$, we will replace for simplicity, as often as possible, the triplet $(d,n,d)$ in all notations by the doublet $(d,n)$ ($\equiv (d,n,d)$). The full notation will still be used for Pearson-Gamma and Pearson-Dirichlet walks whose $q \neq d$.

### 2.4 *The densities of the endpoint position and of the endpoint distance*

For the spherically symmetric distributions considered in the present work, the densities of the endpoint position at $\mathbf{r}$ depend only on $r = \|\mathbf{r}\| > 0$. The pdf's are denoted as:

- $g_{d,n,q}(\mathbf{r}) = g_{d,n,q}(r)$, where $g_{d,n,q}(\mathbf{r})d\mathbf{r}$ represents the probability to find the endpoint of a $PG(d,n,q)$ walk, or the position of the flying particle after $m = nq-1$ Poisson events (section 3.2), within a small volume element $d\mathbf{r}$ at a point $\mathbf{r}$.

- $\psi_{d,n,q}(r)$ represents a radial density, where $\psi_{d,n,q}(r)dr$ is the probability to find the endpoint of a $PG(d,n,q)$ walk at a distance from the starting point ranging between $r$ and $r+dr$.

Similarly, $p_{d,n,q}(\mathbf{r}) = p_{d,n,q}(r)$ and $\omega_{d,n,q}(r)$ ($r<1$) stand for the corresponding pdf's in the case of a $PD(d,n,q)$ walk. The expressions of the previous densities are related through:

$$\begin{cases} \psi_{d,n,q}(r) = \dfrac{2\pi^{d/2}}{\Gamma\left(\dfrac{d}{2}\right)} r^{d-1} g_{d,n,q}(r) \\ \omega_{d,n,q}(r) = \dfrac{2\pi^{d/2}}{\Gamma\left(\dfrac{d}{2}\right)} r^{d-1} p_{d,n,q}(r) \end{cases} \quad (1)$$

Finally, once densities are obtained for a $PD(d,n,q)$ walk of a total length of 1, densities $p_{d,n,q}^{(l)}(\mathbf{r})$ and $\omega_{d,n,q}^{(l)}(r)$ are immediately calculated for an arbitrary total walk length $l$ from:

$$\begin{cases} p_{d,n,q}^{(l)}(\mathbf{r}) = \dfrac{1}{l^d} p_{d,n,q}\left(\dfrac{\mathbf{r}}{l}\right) \\ \omega_{d,n,q}^{(l)}(r) = \dfrac{1}{l^d} \omega_{d,n,q}\left(\dfrac{r}{l}\right) \end{cases} \quad (r < l) \qquad (2)$$

## 3. Generalities about $PG(d,n,q)$ and $PD(d,n,q)$ random walks

### 3.1 Random walk $PG(d,n,q)$

The step lengths of a Pearson-Gamma random walk of $n$ steps in $\mathbb{R}^d$, $PG(d,n,q)$, are i.i.d. gamma distributed, $\gamma(q,1)$, with a shape parameter $q$ and a scale parameter of 1 (appendix A). Then the characteristic function, $\Phi_{d,n,q}(\boldsymbol{\rho}) = \langle e^{i\boldsymbol{\rho}.\mathbf{R}_G} \rangle$, of the endpoint of this walk, $\mathbf{R}_G = \sum_{k=1}^{n} S_k \mathbf{U}_k^{(d)}$, whose distribution is spherically symmetric, is readily obtained from that of a one step walk (eq. 47 of [19], appendix B):

$$\begin{cases} \Phi_{d,1,q}(\rho) = \left\langle e^{i\boldsymbol{\rho}.S_1\mathbf{U}_1^{(d)}} \right\rangle = \dfrac{1}{\Gamma(q)} \int_0^\infty s^{q-1} e^{-s} \Omega_d(s\rho) ds = \\ = \dfrac{2^{(d-2)/2} \Gamma(d/2)}{\rho^q \Gamma(q)} \int_0^\infty x^{q-d/2} J_{(d-2)/2}(x) \exp(-x/\rho) dx \\ \Phi_{d,n,q}(\rho) = \left(\Phi_{d,1,q}(\rho)\right)^n \end{cases} \qquad (3)$$

Consequently (eq. 48 of [19]):

$$\Phi_{d,1,q}(\rho) = \dfrac{1}{(1+\rho^2)^{q/2}} \, {}_2F_1\left(\dfrac{q}{2}, \dfrac{d-q-1}{2}; \dfrac{d}{2}; \dfrac{\rho^2}{(1+\rho^2)}\right) \qquad (4)$$

where $_2F_1(a,b;c;z)$ is a Gauss hypergeometric function. When $q = d-1+\Delta$, with $\Delta = 0,1$, the c.f. $\Phi_{d,n,d-1+\Delta}(\rho)$ simplifies to (eq. 49 of [19]):

$$\Phi_{d,n,d-1+\Delta}(\rho) = \frac{1}{\left(1+\rho^2\right)^{n(d-1+2\Delta)/2}} \tag{5}$$

For $q = d$, the c.f. $\Phi_{d,1,d}(\rho)$ of eq. 4 is indeed a product of $1/\left(1+\rho^2\right)^{d/2}$ by $_2F_1\left(\frac{d}{2}, -\frac{1}{2}; \frac{d}{2}; \frac{\rho^2}{1+\rho^2}\right) = \frac{1}{\left(1+\rho^2\right)^{1/2}}$ (eq. 15.4.6 of [20]). The Fourier inversion of the c.f. given by eq. 5 yields the density of the endpoint $g_{d,n,d-1+\Delta}(r)$ and thus the radial density $\psi_{d,n,d-1+\Delta}(r)$ of a $n$-step Pearson-Gamma walk $PG(d,n,d-1+\Delta)$ in $\mathbb{R}^d$ ($d+\Delta \geq 2$). These densities are expressed in terms of $K_\nu(x)$, a modified Bessel function of the second kind as [19]:

$$\begin{cases} \nu = \left(n(d-1+2\Delta)-d\right)/2 \\ g_{d,n,d-1+\Delta}(r) = \dfrac{r^\nu K_\nu(r)}{2^{\nu+d-1}\pi^{d/2}\Gamma\left(n(d-1+2\Delta)/2\right)} \\ \psi_{d,n,d-1+\Delta}(r) = \dfrac{r^{\nu+d-1} K_\nu(r)}{2^{\nu+d-2}\Gamma(d/2)\Gamma\left(n(d-1+2\Delta)/2\right)} \end{cases} \tag{6}$$

To model the motion of microorganisms on planar surfaces, Stadje [9] investigated a 2D random flight. A microorganism starts at the origin, moves in straight-line paths at constant speed, and changes its direction after exponentially distributed time intervals. Stadje derived the exact pdf of the position $R$ of the microorganism at time $t$ and the conditional pdf of its position at the time of the $n$th turn. As required, $g_{2,n,1}(r)$ (eq. 6) coincides with the density calculated by Stadje. Conolly and Roberts [10] studied the planar motion of a target consisting of a sequence of finite legs traversed at constant speed $V$, with exponentially distributed lengths, and randomly oriented directions with respect to each other. Making $VT = 1$ in their equation 3, we recover the density $\psi_{2,n,1}(r)$ ($d=2$, $\Delta=0$, eq. 6). They considered further the case of a gamma distributed step length with a shape factor $q=2$. $\psi_{2,n,2}(r)$ coincides with the density they derived ($d=2$, $\Delta=1$, eq. 6). To study the relation between the Boltzmann equation and the underlying stochastic processes, Zoia et al. [11] investigated very recently exponential flights in $\mathbb{R}^d$ ($q=1$). In particular, a free propagator was defined as the probability density of finding a particle at position $r$ at the $n$-th collision, for an infinite medium. They calculated the latter for $d=1$, a case which may find applications in the field of electron transport in nanowires or carbon nanotubes. Eq. 50 of [11] coincides with $g_{1,n,1}(r)$ (eq. 6 for $d=1$, $\Delta=1$). In the case $d=2$, which is relevant to the study of the dynamics of chemical and biological species on surfaces, the free propagator (eq. 62 of [11]) coincides with $g_{2,n,1}(r)$ (eq. 6). Finally, the pdf $\psi_{3,5,3}(r) = \dfrac{r^{21/2} K_{17/2}(r)}{92897280\sqrt{2\pi}}$ is compared in figure 2 to that found from Monte-Carlo simulation of $10^8$ Pearson-Gamma random walks of 5 steps in $\mathbb{R}^3$ ($q=d=3$).

### 3.2 Random walk $PD(d,n,q)$

The step lengths of such walks have a Dirichlet distribution $D_m(\boldsymbol{q}^{(n)})$, where $\boldsymbol{q}^{(n)}$ is a $n$-dimensional vector whose components are all equal to $q$ ($q>0$) (appendix A).

#### 3.2.1 $q=1$

$PD(d,n,1)$ may be described as a constrained Pearson-Gamma random walk with exponentially distributed step lengths. Therefore, the step lengths are uniformly distributed over the unit $(n-1)$ simplex ([19] and eq. A-2 with $\alpha_i = q = 1$). The 1D case which was investigated by Franceschetti [12] will not be further considered. Walks $PD(d,n,1)$ are equivalent to random flights performed by a particle which starts from the origin at time $t=0$, moves with a constant and finite velocity $c$ in an initial random direction. It flies until it chooses instantaneously a new direction, at a random time determined by a homogeneous Poisson process, independently of the previous direction [13-17]. For a given time interval $t$, the total length of the flight is then fixed, $l = ct$. The value $q=1$ results from the uniform distribution over the unit $(n-1)$ simplex of the inter-arrival times between successive Poisson events given that the number of events occurring in a unit time interval is $n-1$ [21]. The conditional density of the position of the particle at time $t$, given the number $m=n-1$ of Poisson events that occurred up to $t$, was obtained for any $m$ for $d=2$ and for $d=4$ by Orsingher and De Gregorio [13] and by Kolesnik [14-15]. The densities $p_{d,n,1}(r)$ are readily derived from the latter conditional densities and vice-versa. The density $p_{d,2,1}(r)$ is for instance directly obtained from eq. 2.25 of ref. [13] for any dimension $d \geq 2$:

$$p_{d,2,1}(r) = \frac{2^{d-3}\Gamma(d/2)}{\pi^{d/2}}(1-r^2)^{(d-3)/2} \, _2F_1\left(d-2, \frac{1}{2}; \frac{d}{2}; r^2\right) \quad (r<1) \tag{7}$$

The latter density reduces to an even polynomial in $r$ of degree $d-4$ for even $d \geq 4$. Kolesnik [17] derived explicit expressions of the densities $p_{6,n,1}(r)$ of walks of $n$ steps in 6D which reduce to even polynomials of finite orders. It is for instance, an even polynomial of degree 6 in $r$ for $n=3$ (eq. 15 of [17]).

### 3.2.2 $q=2$ [18]

Beghin and Orsingher [18] studied a planar random motion at finite constant velocity in which a particle changes direction at even-valued Poisson events $(q=2)$. They calculated, among others, the density of the particle position at time $t$ given the number of reorientations between 0 and $t$. Interestingly, this density is a weighted mixture of pdf's of the motion of a particle changing direction at all Poisson events [18]. The density of the endpoint position of the 2D Pearson-Dirichlet walk $PD(2,n,2)$ was indeed proven to be [18]:

$$\begin{cases} p_{2,n}(r) = \sum_{h=0}^{\left\lfloor \frac{n}{2} \right\rfloor} w_{2,n}(h) p_{2,n+2h,1}(r) \\ p_{2,n+2h,1}(r) = \frac{(n+2h-1)}{2\pi}(1-r^2)^{\frac{n-3}{2}+h} \end{cases} \quad (r<1) \tag{8}$$

The weights $w_{2,n}(h)$, whose explicit forms depend on the parity of $n$, are given by eq. 3.3 of [18] and are reproduced in eq. 56 of [19]. They will not be given here as we will derive later in the article a unique expression of the weights $w_{d,n}(h)$ which is valid for any $d \geq 2$ and any $n \geq 2$ (eq. 24).

### 3.2.3 $q = \frac{d}{2} - 1$ $(d \geq 3)$ and $q = d - 1$ $(d \geq 2)$ [19]

The endpoint densities $p_{d,n,q}(r)$ $(r < 1)$ of these two families of Pearson-Dirichlet walks were shown to be (table 1 of [19]):

$$q = \frac{d}{2} - 1, \ k = n(d-2) + 2 \ \begin{cases} \alpha = (n-1)(d-2)/2 \\ p_{d,n,q}(r) = \dfrac{\Gamma(\alpha + d/2)}{\Gamma(\alpha)\pi^{d/2}}(1-r^2)^{\alpha-1} \end{cases} \quad (9)$$

and

$$q = d - 1, \ k = n(d-1) + 1 \ \begin{cases} \beta = (n-1)(d-1)/2 \\ p_{d,n,q}(r) = \dfrac{\Gamma(\beta + d/2)}{\Gamma(\beta)\pi^{d/2}}(1-r^2)^{\beta-1} \end{cases} \quad (10)$$

These densities share a simple geometrical interpretation which is illustrated in figure 4 of [19] for the case of a two-step planar walk $PD(2,n,1)$. The distribution of the endpoint of this 2-step walk on a disc of unit radius, centered at the starting point, is identical to the distribution of the projection on the disc of a point M uniformly distributed over the surface of the 3D unit sphere. More generally, the endpoint distributions of the $PD\left(q, n, \frac{d}{2}-1\right)$ and $PD(d,n,d-1)$ walks are identical to the distributions of the projection in the walk space $\mathbb{R}^d$ of a point randomly chosen on the surface of the unit hypersphere in $\mathbb{R}^k$ where the dimensions $k = k(d,n,q)$ are given in eqs 9 and 10. The densities 9 and 10 are indeed nothing else than eq. B-2 with these values of $k$ and with $j = d$. Such walks were named "hyperspherical uniform" in [19].

Random flights may equally be associated to the previous random walks when $q$ is an integer larger than 1 [18-19]. The conditional pdf of the times of occurrence of Poisson events in (0,1], given their number $N(1) = m = nq - 1$, is $p(t_1, t_2, ..., t_{nq-1} | N(1) = nq - 1) = (nq - 1)! (0 < t_1 < t_2 < ... < t_{nq-1} \leq 1)$ [21] and the inter-arrival times distribution is thus a Dirichlet distribution $D_{nq-1}(\alpha^{(nq)} = (1, 1, ..., 1))$. Amalgamating the $nq$ variables $q$ by $q$ (appendix A) gives the sought-after Dirichlet distribution $D_m(q^{(n)})$. During its flight, a particle changes then its direction at every $q$ Poisson events, $q - 1$ intermediate events being ineffective (figure 1).

## 4. Connection between $PG(d,n,q)$ and $PD(d,n,q)$ random walks $(d \geq 2, n \geq 2)$

The stochastic relation:

$$S \triangleq SL \tag{11}$$

where $S$ is gamma distributed, $\gamma(q_S, 1)$, and $L = (L_1, L_2, ..., L_n)$ is Dirichlet distributed, $D_m(q^{(n)})$, $S$ and $L$ being independent, defines a Gamma-Liouville distribution ([22] p. 148). If $q_S = nq$, then the vector $S = (S_1 \triangleq SL_1, S_2 \triangleq SL_2, ..., S_n \triangleq SL_n)$ has independent $\gamma(q, 1)$ components [22]. Together with independent and random orientations of the $n$ steps, characterized by $n$ i.i.d. unit vectors $U_i^{(d)}$ $(i = 1, ..., n)$, $S$ defines a Pearson-Gamma random walk $PG(d, n, q)$ whose endpoint position is:

$$R_G = \sum_{i=1}^{n} S_i U_i^{(d)} \triangleq S\left(\sum_{i=1}^{n} L_i U_i^{(d)}\right) \tag{12}$$

Similarly $L$ defines a Pearson-Dirichlet random walk $PD(d, n, q)$ whose endpoint position is:

$$R_D = \sum_{i=1}^{n} L_i U_i^{(d)} \tag{13}$$

The connection between the endpoint positions $\mathbf{R}_G$ and $\mathbf{R}_D$ of the random walks $PG(d,n,q)$ and $PD(d,n,q)$, which results from eq. 11, can then be encapsulated in the following stochastic representation:

$$\mathbf{R}_G \triangleq S\mathbf{R}_D \qquad (14)$$

where $S$ and $\mathbf{R}_D$ are independent. The densities $g_{d,n,q}(r=\|\mathbf{r}\|)$ and $\psi_{d,n,q}(r)$ of a $PG(d,n,q)$ random walk are then expressed from eq. 14 in term of the endpoint density of the parent Pearson-Dirichlet walk as:

$$g_{d,n,q}(r) = \frac{1}{\Gamma(nq)} \int_r^\infty s^{nq-1} \exp(-s) p_{d,n,q}^{(s)}(r) \, ds \qquad (15)$$

But, from eq. 2: $p_{d,n,q}^{(s)}(r) = \frac{1}{s^d} p_{d,n,q}\left(\frac{r}{s}\right)$. Then:

$$g_{d,n,q}(r) = \frac{r^{nq-d}}{\Gamma(nq)} \int_1^\infty \exp(-rt) t^{nq-d-1} p_{d,n,q}\left(\frac{1}{t}\right) dt \qquad (16)$$

with $p_{d,n,q}(r)=0$ for $r>1$. The density $g_{d,n,q}(r)$ of the endpoint position of a Pearson-Gamma walk is thus obtained from the Laplace transform of the function $t^{nq-d-1} p_{d,n,q}(1/t)$. Similarly, the density $\psi_{d,n,q}(r)$ of the endpoint distance is given by:

$$\psi_{d,n,q}(r) = \frac{2\pi^{d/2} r^{nq-1}}{\Gamma(d/2)\Gamma(nq)} \int_1^\infty \exp(-rt) t^{nq-d-1} p_{d,n,q}(1/t) \, dt \qquad (17)$$

From integral 3.387.3 of [24], $\int_1^\infty (x^2-1)^{\beta-1} e^{-rx} dx = \frac{\Gamma(\beta)}{\sqrt{\pi}} \left(\frac{2}{r}\right)^{\beta-1/2} K_{\beta-1/2}(r)$, it is readily verified that $p_{d,n,d-1}(r) \propto (1-r^2)^{\beta-1}$ (eq. 10) and $\psi_{d,n,d-1}(r) \propto r^{((n(d-1)+d)/2)-1} K_{\beta-1/2}(r)$ (eq. 6 with $\Delta=0$) are related through eq. 17.

# 5. *Guess of the density of the endpoint position of a $PD(d,n)$ random walk*

The density of the endpoint position of a linear 1D walk, $PD(1,n)$, was shown to be a mixture of a continuous density and of a discrete contribution consisting of two delta peaks placed at the boundaries of the interval [-1,1] [12]. Indeed, a walker has a non-zero probability, $2^{-n}$, to keep walking in the same direction and thus to reach either -1 or +1 in $n$ steps.

As emphasized by Franceschetti, the latter results don't give insights on what happens in higher dimensions. By contrast, the mathematical forms, powers of $(1-r^2)$, of the densities $p_{d,n,q}(r)$ $(n \geq 2)$ are similar for $d=2$ and for $d>2$ for $q=d-1$ (eq. 10) and for all values of $d>3$ for $q=\dfrac{d}{2}-1$ (eq. 9).

As a direct calculation of the inverse Laplace transform of eqs 16 and 17 appeared as out of our reach, we chose another method to solve this problem. From the previous remarks, it is worth investigating the consequences of the assumption that any density $p_{d,n}(r)$ $(q=d>2, n \geq 2)$ of the third family, $q=d$, is too a weighted mixture of powers of $(1-r^2)$, as is the density $p_{2,n}(r)$ $(n \geq 2)$ calculated by Beghin and Orsingher [18] (eq. 8). We assume therefore that:

$$p_{d,n}(r) = \sum_{h=0}^{\left\lfloor \frac{n}{2} \right\rfloor} w_{d,n}(h) f_{d,n,h}(r) \qquad (r<1) \qquad (18)$$

It remains first to guess the densities $f_{d,n,h}(r)$, and from them to guess the weights $w_{d,n}(h)$, and second to prove that the known densities $g_{d,n}(r)$ and $\psi_{d,n}(r)$ (eq. 6 with $\Delta=1$) are retrieved respectively from eqs 16 and 17. Indeed, the uniqueness of the inverse Laplace transform guarantees in that case that the guessed density (eq. 18) is the exact density.

## 5.1 Assumption about the densities $f_{d,n,h}(r), h = 0,..., \left\lfloor \frac{n}{2} \right\rfloor$

We observe that the first term, $h=0$, in the sum which gives $p_{2,n}(r)$ (eq. 8) is the density $p_{2,n,1}(r)$ (eq. 10, $q = d-1 = 1$). When $d \to \infty$, we expect that $p_{d,n,d}(r)$ becomes closer and closer to $p_{d,n,d-1}(r) = \frac{\Gamma(\beta + d/2)}{\Gamma(\beta)\pi^{d/2}}(1-r^2)^{\beta-1}$, where $\beta$ is recalled to be $(n-1)(d-1)/2$. For this reason, we assume that $f_{d,n,0}(r) = p_{d,n,d-1}(r)$. We will show in section 5.2 that the weight $w_{d,n}(0)$ tends consistently to 1 when $d \to \infty$. Monte-Carlo simulations of $PD(d,n)$ walks, performed for $n = 2, 3, 4$ and $d = 2, 3, 4$, were found to be consistent with eq. 18 with the additional assumption that $f_{d,n,h}(r) = \frac{\Gamma(\beta+h+d/2)}{\Gamma(\beta+h)\pi^{d/2}}(1-r^2)^{\beta+h-1}$, in agreement with eq. 8 for $d = 2$. Our starting assumption is then that the pdf of endpoint position of the Pearson-Dirichlet walk $PD(d,n)$ is a weighted mixture of hyperspherical random walks (section 3.2.3) with hyperspace dimensions $k_{d,n}(h) = n(d-1) + 1 + 2h$:

$$p_{d,n}(r) = \sum_{h=0}^{\left\lfloor \frac{n}{2} \right\rfloor} w_{d,n}(h) \left[ \frac{\Gamma\left(\beta + \frac{d}{2} + h\right)}{\Gamma(\beta+h)\pi^{d/2}}(1-r^2)^{\beta-1+h} \right] \quad (r<1) \quad (19)$$

## 5.2 Assumption about the weights $w_{d,n}(h)$

The moments of the endpoint distance of a $PG(d,n)$ walk, $\langle R_G^{2x} \rangle = \int_0^\infty r^{2x} \psi_{d,n}(r)\, dr$ $(x \geq 0)$ are obtained on the one hand by a direct calculation of this integral from the known density $\psi_{d,n}(r)$ (eq. 6), $\langle R_G^{2x} \rangle = 4^x \left(\frac{d}{2}\right)_x \left(\frac{n(d+1)}{2}\right)_x$ (eq. C-13, appendix C) and on the other hand by a calculation from the right member of eq. 17 in which $p_{d,n}(r)$ is replaced by its expression (eq. 19).

The following classical integral (3.251.3 of [24]),

$$\int_1^\infty t^{-n(d-1)-2h-2x}\left(t^2-1\right)^{\beta+h-1} dt = \frac{\Gamma\left(\frac{d}{2}+x\right)\Gamma(\beta+h)}{2\Gamma\left(\frac{n(d-1)+1}{2}+x+h\right)}$$

, gives finally:

$$\sum_{h=0}^{\lfloor\frac{n}{2}\rfloor} \frac{w_{d,n}(h)}{\left(\frac{n(d-1)+1}{2}+h\right)_x} = \frac{4^x\left(\frac{n(d+1)}{2}\right)_x}{(nd)_{2x}} \quad (x \geq 0) \quad (20)$$

Eq. 20 was previously written for $d=2$ (see eq. 57 of [19]). Eq. 20 is actually an identity which can be used further to obtain explicit expressions of the $\left\lfloor\frac{n}{2}\right\rfloor$ unknown weights $w_{d,n}(h)$ from sets of linear equations for moderate values of $n$. These sets of linear equations consist for instance of the normalization relation $\left(\sum_{h=0}^{\lfloor\frac{n}{2}\rfloor} w_{d,n}(h)=1\right)$ and of $\left\lfloor\frac{n}{2}\right\rfloor-1$ equations obtained for convenient values of $x>0$. Then, these sets of equations were solved with Maple. An example is given below for $n=5$:

$$\begin{cases} w_{d,5}(0)+w_{d,5}(1)+w_{d,5}(2)=1 \\ \\ \dfrac{w_{d,5}(0)}{5d-4}+\dfrac{w_{d,5}(1)}{5d-2}+\dfrac{w_{d,5}(2)}{5d}=\dfrac{d+1}{d(5d+1)} \\ \\ \dfrac{w_{d,5}(0)}{(5d-4)(5d-2)}+\dfrac{w_{d,5}(1)}{5d(5d-2)}+\dfrac{w_{d,5}(2)}{5d(5d+2)}=\dfrac{(d+1)(5d+7)}{d(5d+1)(5d+2)(5d+3)} \end{cases} \quad (21)$$

where the three equations are obtained from eq. 20 respectively for $x=0,1,2$. The solution is readily verified to be that given in table 1. As required, the weights $w_{2,2}(h)$, $w_{2,3}(h)$ and $w_{2,4}(h)$, calculated for $d=2$ from the expressions given in table 1, are identical with those reported in table 2 of [18]. The weights, which were calculated with the previous method for $2 \leq n \leq 9$ (some of them are given in table 1), can be written as products of two factors which depend both on $n$, $w_{d,n}(h) = B(n,h)c_{d,n}(h)$, but the first factor, $B(n,h)$, is independent of $d$. The following expression of the factor $c_{d,n}(h)$ is deduced from the latter calculations:

$$\begin{cases} c_{d,n}(h) = \dfrac{\prod_{j=1}^{\lfloor \frac{n}{2} \rfloor - h} (nd - (n+1) + 2j + 2h)}{\prod_{j=1}^{\lfloor \frac{n}{2} \rfloor} (n(d+1) - 2j)}, & h = 0, \ldots, \left\lfloor \dfrac{n}{2} \right\rfloor - 1 \\ c_{d,n}\left(\left\lfloor \dfrac{n}{2} \right\rfloor\right) = \dfrac{1}{\prod_{j=1}^{\lfloor \frac{n}{2} \rfloor} (n(d+1) - 2j)} \end{cases} \quad (22)$$

An alternative writing of the latter factor is given below. The factors $B(n,h)$ $\left(h = 0, \ldots, \left\lfloor \dfrac{n}{2} \right\rfloor\right)$, which are such that $B(n,0) = 1$, $B(2,0) = B(2,1) = 1$ and $B(3,1) = 3$, are given in table 2 for $4 \leq n \leq 9$. Using the On-Line Encyclopedia of Integer Sequences [25], these numbers are recognized to be Bessel numbers (see also [26]). Finally:

$$\begin{cases} B(n,h) = \dfrac{n!}{h!(n-2h)!2^h} = \dfrac{2^h \left(-\dfrac{n}{2}\right)_h \left(-\dfrac{n}{2} + \dfrac{1}{2}\right)_h}{h!} \\ c_{d,n}(h) = \dfrac{\sqrt{\pi} \times \Gamma(nd)}{2^{nd+h-1} \Gamma\left(\dfrac{n(d+1)}{2}\right) \Gamma\left(\dfrac{n(d-1)+1}{2} + h\right)} \end{cases} \quad h = 0, \ldots, \left\lfloor \dfrac{n}{2} \right\rfloor \quad (23)$$

In sum:

$$w_{d,n}(h) = \dfrac{\sqrt{\pi} B(n,h) \Gamma(nd)}{2^{nd+h-1} \Gamma\left(\dfrac{n(d+1)}{2}\right) \Gamma\left(\dfrac{n(d-1)+1}{2} + h\right)} \quad h = 0, \ldots, \left\lfloor \dfrac{n}{2} \right\rfloor \quad (24)$$

All weights of table 1 are readily retrieved from eq. 24. The full expression of the pdf of the endpoint $p_{d,n}(r)$ $(d, n \geq 2)$ is then assumed to be:

$$p_{d,n}(r) = \frac{\Gamma(nd)}{2^{nd-1}\pi^{(d-1)/2}\Gamma\left(\frac{n(d+1)}{2}\right)} \left\{ \sum_{h=0}^{\lfloor n/2 \rfloor} \frac{B(n,h)}{2^h \Gamma(\beta+h)} (1-r^2)^{\beta+h-1} \right\} \qquad (r<1) \qquad (25)$$

and consequently (eq. 1):

$$\omega_{d,n}(r) = \frac{\sqrt{\pi}\,\Gamma(nd)\,r^{d-1}}{2^{nd-2}\Gamma\left(\frac{d}{2}\right)\Gamma\left(\frac{n(d+1)}{2}\right)} \left\{ \sum_{h=0}^{\lfloor n/2 \rfloor} \frac{B(n,h)}{2^h \Gamma(\beta+h)} (1-r^2)^{\beta+h-1} \right\} \qquad (r<1) \qquad (26)$$

The density $p_{d,n}(r)$ reduces to an even polynomial in $r$ when $2\beta = (n-1)(d-1)$ is even. Expressing the Bessel number in term of Pochhammer symbols (eq. 23), the bracketed sum in eqs 25 and 26 can be expressed in term of a product of a power of $(1-r^2)$ and a Gauss hypergeometric function whose series terminates in all cases because either $-\frac{n}{2}$ or $-\frac{n}{2}+\frac{1}{2}$ is a negative integer according to whether $n$ is even or odd. A form of $p_{d,n}(r)$ (and consequently that of $\omega_{d,n}(r)$ from eq. 1), equivalent to that of eq. 25, is thus given by:

$$\begin{cases} p_{d,n}(r) = a_{d,n}(1-r^2)^{\beta-1}\, {}_2F_1\left(-\frac{n}{2},\ -\frac{n}{2}+\frac{1}{2}\ ;\ \beta\ ;\ 1-r^2\right) \\ \\ a_{d,n} = \dfrac{\Gamma(nd)}{2^{nd-1}\pi^{(d-1)/2}\Gamma\left(\frac{n(d+1)}{2}\right)\Gamma(\beta)} \end{cases} \qquad (r<1) \qquad (27)$$

When $d$ tends to infinity for a given $n$, the weights $w_{d,n}(h)$ (eq. 24) are such that $\lim_{d\to\infty} w_{d,n}(h) = 0$ for $h > 0$ while $\lim_{d\to\infty} w_{d,n}(0) = 1$ as discussed in section 5.1 (see also table 1 for $2 \le n \le 5$). Eq. 27 shows consistently that the density $p_{d,n}(r)$ tends to coincide with $p_{d,n,d-1}(r) \propto (1-r^2)^{\beta-1}$ (eq. 10) when $d \to \infty$, for $n$ fixed, as the $d$-dependent factors, which vary then as $d^h$ $\left(h=1,...,\left\lfloor\frac{n}{2}\right\rfloor\right)$, appear solely at the denominators of the $\left\lfloor\frac{n}{2}\right\rfloor$ terms of the finite hypergeometric series. The first term $(h=0)$ is equal to 1.

The method based on the generation of i.i.d. gamma random variables is the simplest method for simulating Dirichlet distributions that was used in the present work for performing Monte-Carlo simulations of Pearson-Dirichlet random walks. Examples of such simulations are shown in figure 3 for a $PD(3,5)$ walk and in figure 4 for two $PD(4,n)$ with $n=3,4$ steps where they are compared to the guessed theoretical densities (eq. 26). The deviations between the simulated and the guessed densities are typically of the order of or less than the line thicknesses.

When we insert the guessed weights $w_{d,n}(h)$ (eq. 24) into the left member of eq. 20, we obtain:

$$= \frac{\sqrt{\pi}\,\Gamma(nd)}{2^{nd-1}\Gamma\left(\frac{n(d+1)}{2}\right)\Gamma\left(\frac{n(d-1)+1}{2}+x\right)} \, {}_2F_1\left(-\frac{n}{2}, -\frac{n}{2}+\frac{1}{2}; \frac{n(d-1)+1}{2}+x\,;\,1\right) =$$

$$= \frac{\sqrt{\pi}\,\Gamma(nd)\,\Gamma\left(\frac{n(d+1)}{2}+x\right)}{2^{nd-1}\Gamma\left(\frac{nd}{2}+x\right)\Gamma\left(\frac{nd}{2}+\frac{1}{2}+x\right)\Gamma\left(\frac{n(d+1)}{2}\right)} \tag{28}$$

using the relation, ${}_2F_1(\alpha,\beta;\gamma;1)=\dfrac{\Gamma(\gamma)\Gamma(\gamma-\alpha-\beta)}{\Gamma(\gamma-\alpha)\Gamma(\gamma-\beta)}$ (eq. 9.122.1 of [24]) when $\gamma>\alpha+\beta$, a condition which is fulfilled by the parameters of ${}_2F_1(\,,\,;\,;1)$ of eq. 28. We finally apply the duplication formula of gamma functions, $\Gamma\left(\frac{nd}{2}+x\right)\Gamma\left(\frac{nd}{2}+\frac{1}{2}+x\right)=2^{1-nd-2x}\sqrt{\pi}\,\Gamma(nd+2x)$, to get the right member of eq. 20:

$$\frac{\sqrt{\pi}\,\Gamma(nd)\,\Gamma\left(\frac{n(d+1)}{2}+x\right)}{2^{nd-1}\Gamma\left(\frac{nd}{2}+x\right)\Gamma\left(\frac{nd}{2}+\frac{1}{2}+x\right)\Gamma\left(\frac{n(d+1)}{2}\right)} = \frac{4^x\left(\frac{n(d+1)}{2}\right)_x}{(nd)_{2x}} \tag{29}$$

Therefore, the weights $w_{d,n}(h)$, whose exact values were obtained directly from eq. 20 only for $2 \le n \le 9$, are seen to yield the right member of this equation for any value of $x \ge 0$, $d \ge 2$ and $n \ge 2$. The very existence of such solutions is already a strong indication of the correctness of eqs 25 and 26. Additional evidence is provided in appendix C through the direct calculation of the moments $\langle R_D^2 \rangle$ and $\langle R_D^4 \rangle$ on the one hand from the Dirichlet distribution without recourse to the explicit density and on the other hand from the explicit density (eq. 26). Further, it is shown in appendix C that the moments $\langle R_G^k \rangle$ are equal to $\langle S^k \rangle \langle R_D^k \rangle$, for any $k \geq 0$, in agreement with the stochastic representation eq. 14. The latter result might be considered as an indirect proof of the Laplace transform of eq. 17 as the density $\psi_{d,n}(r)$ (eq. 6) is uniquely determined by its moments (appendix C). We prefer however to prove directly below that the Laplace transform of $t^{n(d-1)-1} p_{d,n}(1/t)$ (eq. 17), together with the guessed density $p_{d,n}(r)$ (eq. 25), yields $\psi_{d,n}(r)$.

## 6. A proof of eq. 17 from $p_{d,n}(r)$ (eq. 25)

To simplify the writing of the next equations, we define the following parameters :

$$\begin{cases} n = 2p + \varepsilon \ (\varepsilon = 0,1), \ p = \left\lfloor \dfrac{n}{2} \right\rfloor \\ \alpha = (n-1)d + \varepsilon \\ \mu = \dfrac{n(d+1) - d}{2} \end{cases} \quad (30)$$

The density $\psi_{d,n}(r) = \dfrac{r^{\mu+d-1} K_\mu(r)}{2^{\mu+d-2} \Gamma(d/2) \Gamma(n(d+1)/2)}$ (eq. 6) is related to the exact density $p_{d,n}(r)$ by a Laplace transform (eq. 17). We must then prove that:

$$\left(\frac{2}{r}\right)^{\beta - 1/2} K_\mu(r) = \sqrt{\pi} \left\{ \sum_{h=0}^{p} \frac{B(n,h)}{2^h \Gamma(\beta + h)} \int_1^\infty \exp(-rt) t^{n-2h} (t^2 - 1)^{\beta + h - 1} dt \right\} \quad (31)$$

The following integral:

$$\int_1^\infty \exp(-rx) x^{2s+\varepsilon} (x^2 - 1)^{\nu - 1} dx = \frac{\Gamma(\nu)}{\sqrt{\pi}} \left(\frac{2}{r}\right)^{\nu - 1/2} \left\{ \sum_{k=0}^{s} \binom{s}{k} (\nu)_{s-k} \left(\frac{2}{r}\right)^{s-k} K_{\nu + s - \frac{1}{2} + \varepsilon - k}(r) \right\} \quad (32)$$

, where $s \geq 0$ is an integer and $\varepsilon = 0, 1$, follows readily from integrals 3.387.3 ($s=0, \varepsilon=0$) and 3.389.4 ($s=0, \varepsilon=1$) of ref. [24] as $x^{2s+\varepsilon} = x^{\varepsilon}\left((x^2-1)+1\right)^s$. For $v = \beta + h$, eq. 31 reduces then to:

$$K_\mu(r) = \sum_{h=0}^{p} \frac{B(n,h)}{r^h} \left\{ \sum_{k=0}^{p-h} \binom{p-h}{k} (\beta+h)_{p-h-k} \left(\frac{2}{r}\right)^{p-h-k} K_{\beta+p-\frac{1}{2}+\varepsilon-k}(r) \right\} \quad (33)$$

As:

$$\begin{cases} (\beta+h)_{p-h-k} = \frac{\Gamma(\beta+p-k)}{\Gamma(\beta)} \times \frac{\Gamma(\beta)}{\Gamma(\beta+h)} = \frac{(\beta)_{p-k}}{(\beta)_h} \\ \beta + p - \frac{1}{2} + \varepsilon - k = \alpha - k \end{cases} \quad (34)$$

we obtain, after interchanging the order of the summations over $h$ and $k$:

$$K_\mu(r) = \sum_{k=0}^{p} (\beta)_{p-k} \left\{ \sum_{h=0}^{p-k} \frac{B(n,h)}{2^h (\beta)_h} \binom{p-h}{k} \right\} \left(\frac{2}{r}\right)^{p-k} K_{\alpha-k}(r) \quad (35)$$

We must show next, using eq. D-2 (appendix D), that eq. 35 reduces to:

$$K_\mu(r) = \sum_{k=0}^{p} \binom{p}{k} (\alpha)_{p-k} \left(\frac{2}{r}\right)^{p-k} K_{\alpha-k}(r) \quad (36)$$

Comparing eqs 35 and 36, the proof that eq. 33 holds will finally follow from that of:

$$\binom{p}{k} (\alpha)_{p-k} = (\beta)_{p-k} \sum_{h=0}^{p-k} \binom{p-h}{k} \frac{B(n,h)}{2^h (\beta)_h} \quad (37)$$

The falling factorial is classically defined as $(a)_{(n)} = \frac{\Gamma(a+1)}{\Gamma(a-n+1)} = (-1)^n (-a)_n$ and thus $\binom{p}{k} = \frac{(-1)^k (-p)_k}{k!}$. Therefore:

$$\binom{p-h}{k} = \frac{\Gamma(p-k+1)}{k!\,\Gamma(p-h-k+1)} \times \frac{\Gamma(p-h+1)}{\Gamma(p+1)} \times \frac{\Gamma(p+1)}{\Gamma(p-k+1)}$$

$$= \frac{(p-k)_{(h)}(p)_{(k)}}{k!\,(p)_{(h)}} = (-1)^k \frac{(k-p)_h(-p)_k}{k!\,(-p)_h} = \binom{p}{k}\frac{(k-p)_h}{\left(-\dfrac{n}{2}+\dfrac{\varepsilon}{2}\right)_h} \tag{38}$$

From the expression of Bessel numbers in term of Pochhammer symbols (eq. 23) and from eq. 38, eq. 37 simplifies to:

$$(\alpha)_{p-k} = (\beta)_{p-k} \sum_{h=0}^{p-k} \frac{\left(-\dfrac{n}{2}+\dfrac{1-\varepsilon}{2}\right)_h (k-p)_h}{(\beta)_h\, h!} \tag{39}$$

Eq. 39 can be rewritten in term of a Gauss hypergeometric function as the sum terminates because the factor $(k-p)_h$ is zero for $h > p-k$:

$$(\alpha)_{p-k} = (\beta)_{p-k}\, {}_2F_1\left(-\frac{n}{2}+\frac{1-\varepsilon}{2}, k-p\,;\, \beta\,;1\right) \tag{40}$$

The value of ${}_2F_1(\,,\,;\,;\,1)$ (eq. 9.122 of [24]), once plugged into eq. 40, completes the proof:

$${}_2F_1\left(-\frac{n}{2}+\frac{1-\varepsilon}{2}, k-p\,;\, \beta\,;1\right) = \frac{\Gamma(\beta)\Gamma(\alpha+p-k)}{\Gamma(\beta+p-k)\Gamma(\alpha)} = \frac{(\alpha)_{p-k}}{(\beta)_{p-k}} \tag{41}$$

The Laplace transform given by eq. 17 is then proven to relate, as required, the pdf $\psi_{d,n}(r)$ (eq. 6) to $p_{d,n}(r)$ (eq. 25), alternatively expressed as $p_{d,n}(r) = \sum_{h=0}^{p} w_{d,n}(h) f_{d,n,h}(r)$ (eq. 19). The latter result proves ipso facto that the pdf $g_{d,n}(r)$ is related to $p_{d,n}(r)$ by the Laplace transform of eq. 16. To conclude, the uniqueness of the inverse Laplace transform ensures that $p_{d,n}(r)$ (eq. 25) is the exact density of the endpoint position of the Pearson-Dirichlet random walk $PD(d,n)$.

Another option consists in deriving the characteristic function of $\boldsymbol{R}_G$ from that of $\boldsymbol{R}_D$ and in proving that it is identical with the characteristic function calculated directly for a $PG(d,n)$ random walk (eq. 5). This calculation is presented in appendix E. The characteristic function $\Phi_{d,n}^D(\rho = \|\boldsymbol{\rho}\|) = \langle e^{i\boldsymbol{\rho}.\boldsymbol{R}_D}\rangle$ of $\boldsymbol{R}_D$ is a weighted sum of c.f.'s of uniformly oriented unit vectors of $\mathbb{R}^{k(h)}$, where $k(h) = n(d-1)+1+2h\left(h=0,...,\left\lfloor\frac{n}{2}\right\rfloor\right)$. The c.f. is indeed given by:

$$\Phi_{d,n}^D(\rho) = \sum_{h=0}^{\left\lfloor\frac{n}{2}\right\rfloor} w_{d,n}(h)\Omega_{n(d-1)+1+2h}(\rho) \tag{42}$$

It is explicitly given in appendix E (eq. E-2). Appendix E yields in addition an inverse Laplace transform (eq. E-11) that we failed to find in the literature.

When $q$ is a positive integer, the conditional densities of particle positions $f_{d,n,q}(\boldsymbol{x}|m=nq-1)$, given the number of Poisson events $m=nq-1$, of the random flights associated with the Pearson-Dirichlet random walks $PD(d,n,q)$ (section 3.2 and figure 1), are obtained from eq. 2. They are simply expressed as $f_{d,n,q}(\boldsymbol{x}|m=nq-1) = p_{d,n,q}(\boldsymbol{x}/ct)/(ct)^d$, where $p_{d,n,q}(\boldsymbol{x}/ct)$ is given by eq. 9 for $q=\frac{d}{2}-1$ with $d$ even, by eq. 10 for $q=d-1$ and by eq. 19 or eq. 25 for $q=d$.

## 7. Conclusions

The endpoint positions $\boldsymbol{R}_G$ of three families of Pearson-Gamma random walks, $PG(d,n,q)$ with $q=\frac{d}{2}-1$, $d-1$, $d$, have densities with compact closed-forms expressions which are products of powers and modified Bessel functions of the second kind (eq. 6). The associated Pearson-Dirichlet walks $PD(d,n,q)$, with the same parameters $(d,n,q)$, have endpoint positions $\boldsymbol{R}_D$ which are simply related to $\boldsymbol{R}_G$ by the stochastic representation $\boldsymbol{R}_G \triangleq S\boldsymbol{R}_D$, where $S$ is gamma distributed, $\gamma(nq,1)$, and independent of $\boldsymbol{R}_D$. The resulting densities $p_{d,n,q}(\boldsymbol{r})$ of $\boldsymbol{R}_D$ have too simple closed-form expressions which hold for any $d \geq d_0$ and any $n \geq 2$, where $d_0 = 3, 2, 2$ for $q = \frac{d}{2} - 1, d-1, d$ respectively ([19] and present work):

$$p_{d,n,q}(\boldsymbol{r}) = \begin{cases} \dfrac{\Gamma(\alpha + d/2)}{\Gamma(\alpha)\pi^{d/2}} (1-r^2)^{\alpha-1} & \text{for} \quad q = \dfrac{d}{2}-1 \\ \dfrac{\Gamma(\beta + d/2)}{\Gamma(\beta)\pi^{d/2}} (1-r^2)^{\beta-1} & \text{for} \quad q = d-1 \\ \displaystyle\sum_{h=0}^{\lfloor n/2 \rfloor} w_{d,n}(h) \left[ \dfrac{\Gamma(\beta + d/2 + h)}{\Gamma(\beta + h)\pi^{d/2}} (1-r^2)^{\beta-1+h} \right] & \text{for} \quad q = d \end{cases} \quad (43)$$

where $r = \|\boldsymbol{r}\| < 1$, $\alpha = (n-1)(d-2)/2$ and $\beta = (n-1)(d-1)/2$ while $w_{d,n}(h)$ is given by eqs 23 and 24. Equivalently $p_{d,n,d}(\boldsymbol{r})$ can be expressed in term of a Gauss hypergeometric function as:

$$p_{d,n,d}(\boldsymbol{r}) = \frac{\Gamma(nd)}{2^{nd-1}\pi^{(d-1)/2}\Gamma\left(\dfrac{n(d+1)}{2}\right)\Gamma(\beta)} (1-r^2)^{\beta-1} \, {}_2F_1\left(-\frac{n}{2}, -\frac{n}{2}+\frac{1}{2} ; \beta ; 1-r^2\right) \quad (44)$$

The first two densities are those of particular walks, named "hyperspherical uniform" walks in [19], whose endpoint distributions are identical with the distribution of the projection in the walk space $\mathbb{R}^d$ of a point randomly chosen on the surface of the unit hypersphere in a "hyperspace" $\mathbb{R}^k$. The hyperspace dimension is $k = n(d-2)+2$ for $q = \dfrac{d}{2}-1$ and $k = n(d-1)+1$ for $q = d-1$ [19]. The third is a weighted mixture of $1 + \left\lfloor \dfrac{n}{2} \right\rfloor$ hyperspherical uniform walks issued from the second family with $k(h) = n(d-1)+1+2h$ $\left(h = 0, ..., \left\lfloor \dfrac{n}{2} \right\rfloor\right)$. The densities $\omega_{d,n,d}(r)$ of these three families of Pearson-Dirichlet random walks are related through eq. 1 to the corresponding densities given by eqs 43 and 44.

## *Acknowledgements*

I wish to thank an anonymous Referee for his constructive suggestions for improvement.

# *Appendix A: the Gamma and the Dirichlet distributions*

A random variable $X$ is gamma distributed, with a shape parameter $q > 0$ and a scale parameter $\theta$, if its probability density function $p_G(x)$ is [22-23]:

$$p_G(x) = \frac{x^{q-1}\exp(-x/\theta)}{\theta^q \Gamma(q)} \quad (x > 0) \tag{A-1}$$

where $\Gamma(q)$ is the Euler gamma function. The latter gamma distribution will be denoted for brevity as $\gamma(q,\theta)$. The characteristic function of $X$ is $\varphi_X(t) = \langle e^{itX} \rangle = 1/(1-i\theta t)^q$ and the moments $\langle X^k \rangle$ are equal to $\theta^k (q)_k$. A sum $S$ of $K$ independent gamma random variables, $\gamma(\alpha_i, \theta)$ $(i=1,..,K)$, with identical scale parameters and a priori different shape parameters, is a gamma random variable $\gamma\left(\alpha = \sum_{i=1}^{K}\alpha_i, \theta\right)$ as deduced from its c.f. $\varphi_S(t) = \langle e^{itS} \rangle = 1/\left(\prod_{k=1}^{K}(1-it\theta)^{\alpha_k}\right)$. As the scale parameter is irrelevant in the present context, its value will be fixed at 1 from now on.

The Dirichlet distribution can be defined as follows [22]:

- First, consider a set of $n = m+1$ independent random variables, $S_i$ $(>0, \ i=1,..,n)$, with gamma distributions $\gamma(\alpha_i, 1)$ $(i=1,...,n)$
- Second, define $S = \sum_{j=1}^{n} S_j$ and $L_j = S_j/S$ $(j=1,..,n)$ $\left(\sum_{j=1}^{n} L_j = 1\right)$

Then the distribution of $\boldsymbol{L}^{(m)} = (L_1, L_2,..., L_m)$ is a Dirichlet distribution $D_m(\boldsymbol{\alpha}^{(n)})$ with parameters $\boldsymbol{\alpha}^{(n)} = (\alpha_1,...,\alpha_m,\alpha_n)$. Its probability density function is ([22], p. 17):

$$\begin{cases} p_D^{(n)}(l_1,...,l_m) = K_n(\boldsymbol{\alpha}^{(n)}) \prod_{i=1}^{n} l_i^{\alpha_i - 1} \\ l_n = 1 - \sum_{i=1}^{m} l_i, \quad l_i > 0, \ i = 1,..,n \end{cases} \tag{A-2}$$

where $K_n(\boldsymbol{\alpha}^{(n)}) = \Gamma(\alpha) \big/ \left( \prod_{i=1}^{n} \Gamma(\alpha_i) \right)$, $\alpha = \sum_{i=1}^{n} \alpha_i$. Defining a vector $\boldsymbol{\delta}^{(n)} = (\delta_1, ..., \delta_m, 0)$ and $\delta = \sum_{i=1}^{n} \delta_i$, the moment $M_\delta = \left\langle \prod_{i=1}^{m} L_i^{\delta_i} \right\rangle$ is given by [22]:

$$M_\delta = \left\langle \prod_{i=1}^{m} L_i^{\delta_i} \right\rangle = \left( \prod_{i=1}^{m} (\alpha_i)_{\delta_i} \right) \Big/ (\alpha)_\delta \qquad (\text{A-3})$$

If the $n$ components $L_1, ..., L_n \left( \sum_{i=1}^{n} L_i = 1 \right)$ of a vector, whose distribution is $D_m(\boldsymbol{\alpha}^{(n)})$, are grouped into $k$ components $O_1, ..., O_k \left( \sum_{i=1}^{k} O_i = 1 \right)$, then the distribution of $(O_1, ..., O_{k-1})$ is $D_{k-1}(\boldsymbol{\alpha}^{*(k)})$ where each $\alpha_i^* (i = 1, ..., k)$ is the sum of the $\alpha_j$'s corresponding to the components of the initial vector which add up to $O_i$. The previous amalgamation property [22] results directly from the property of the sum of independent gamma random variables described above.

The step lengths of the Pearson-Dirichlet walks we consider in the present work, $(L_1, L_2, ..., L_m)$, have all a Dirichlet distribution $D_m(\boldsymbol{q}^{(n)})$, where $\boldsymbol{q}^{(n)}$ is a $n$-dimensional vector whose components are all equal to $q$ $(q > 0)$. In most cases, $q = \dfrac{d}{2} - 1$, $d - 1$ or $d$ where $d$ is the dimension of the walk space.

# *Appendix B:*
# *A vector uniformly distributed over the surface*
# *of the unit hypersphere in $\mathbb{R}^k$*

The characteristic function $\Omega_k(\boldsymbol{\rho}) = \langle e^{i\boldsymbol{\rho}.\boldsymbol{U}^{(k)}} \rangle$ of $\boldsymbol{U}^{(k)}$, a uniformly oriented unit vector of $\mathbb{R}^k$, depends solely on the modulus $\rho = \|\boldsymbol{\rho}\|$ [22]:

$$\Omega_k(\rho) = \frac{\Gamma(k/2)}{\sqrt{\pi}\,\Gamma((k-1)/2)} \int_0^\pi e^{i\rho\cos(\theta)} \sin^{k-2}(\theta)\,d\theta = \frac{2^{(k-2)/2}\,\Gamma(k/2)}{\rho^{(k-2)/2}} J_{(k-2)/2}(\rho) \qquad \text{(B-1)}$$

where $J_u(\rho)$ is a Bessel function of the first kind.

The joint distribution of any number $j$ of components of $\boldsymbol{U}^{(k)}$ can be obtained by using the amalgamation property of the Dirichlet distribution of $(L_1 = U_1^2,...,L_k = U_k^2)$, whose parameters are all equal to $1/2$ (see for instance [22] or appendix of [19]). It is given by :

$$p_j(u_1,u_2,...,u_j) = \frac{\Gamma(k/2)}{\Gamma((k-j)/2)\,\pi^{j/2}} \left(1 - \sum_{i=1}^j u_i^2\right)^{(k-j-2)/2} \qquad \left(\sum_{i=1}^j u_i^2 < 1\right) \qquad \text{(B-2)}$$

The previous Dirichlet distribution with $\boldsymbol{\alpha}^{(k)} = (1/2,..,1/2)$ and $\boldsymbol{\beta}^{(k)} = (p,0,...,0)$ yields the even moments $\langle U_i^{2p} \rangle$ of a single component of $\boldsymbol{U}^{(k)}$ :

$$\langle U_i^{2p} \rangle = \frac{(1/2)_p}{(k/2)_p} = \frac{(2p-1)!!}{\prod_{j=1}^p (k+2j-2)} \qquad \text{(B-3)}$$

# Appendix C: the moments $\langle R_D^k \rangle$ of a $PD(d,n)$ random walk

The position of the endpoint of a Pearson-Dirichlet walk $PD(d,n)$ is given by a vector of $\mathbb{R}^d$, $\boldsymbol{R}_D = (R_D(1), R_D(2),..., R_D(d))$, which reads $(d \geq 2, n \geq 2)$:

$$\boldsymbol{R}_D = \sum_{i=1}^{n} L_i \boldsymbol{U}_i^{(d)} \qquad (C\text{-}1)$$

where the $\boldsymbol{U}_i^{(d)} = (U_i(1),...,U_i(d)), (i=1,..,n)$, are $n$ independent random vectors whose tips are uniformly distributed over the surface of a unit hypersphere in $\mathbb{R}^d$. Further, the $L_i$'s have a Dirichlet distribution (appendix A) whose parameters are all equal to $d$. The moments $\langle R_D^2 \rangle$ and $\langle R_D^4 \rangle$ of the distance $R_D = \| \boldsymbol{R}_D \|$ can be calculated on the one hand directly from eq. C-1 without additional assumptions and on the other hand from the explicit expression of the distance density $\omega_{d,n}(r)$ (eq. 26).

## C-1) Moments calculated from eq. C-1

The moment $\langle R_D^2 \rangle$ is just $d$ times the moment of the square, $\langle R_D^2(k) \rangle$, of any of the $d$ components of $\boldsymbol{R}_D$ (eq. 12 of [19] with $q=d$), that is:

$$\langle R_D^2 \rangle = d \langle R_D^2(1) \rangle = \frac{d+1}{nd+1} \qquad (C\text{-}2).$$

The fourth-order moment, is obtained from:

$$\langle R_D^4 \rangle = \left\langle \left( \sum_{k=1}^{d} R_D^2(k) \right)^2 \right\rangle = d \langle R_D^4(1) \rangle + d(d-1) \langle R_D^2(1) R_D^2(2) \rangle \qquad (C\text{-}3)$$

where $\langle R_D^4(1) \rangle$ is obtained from eq. 12 of [19] with $q=d$:

$$d \langle R_D^4(1) \rangle = \frac{3(d+1)(n(d+1)+2)}{(nd+1)(nd+2)(nd+3)} \qquad (C\text{-}4)$$

In addition:

$$\langle R_D^2(1) R_D^2(2) \rangle = n(n-1) \langle L_1^2 L_2^2 \rangle \langle U_1^2(1) \rangle^2 + n \langle L_1^4 \rangle \langle U_1^2(1) U_1^2(2) \rangle \qquad (C\text{-}5)$$

The mean $\langle U_1^2(1)\rangle$ is simply $\dfrac{1}{d}$ while $\langle U_1^2(1)U_1^2(2)\rangle = \dfrac{1}{d(d+2)}$ is deduced from:

$$\left\langle \left(\sum_{j=1}^{d} U_1^2(j)\right)^2 \right\rangle = 1 = d\langle U_1^4(1)\rangle + d(d-1)\langle U_1^2(1)U_1^2(2)\rangle \tag{C-6}$$

because $\langle U_1^4(1)\rangle = \dfrac{3}{d(d+2)}$ (eq. B-3). Further, the moments $\langle L_1^2 L_2^2\rangle = \dfrac{(d)_2^2}{(nd)_4}$ and $\langle L_1^4\rangle = \dfrac{(d)_4}{(nd)_4}$ are obtained from eq. A-3. Collecting all terms from eqs C-3 to C-6 gives finally:

$$\langle R_D^4 \rangle = \dfrac{(d+1)(d+2)(n(d+1)+2)}{(nd+1)(nd+2)(nd+3)} \tag{C-7}$$

## C-2) Moments $\langle R_D^k \rangle$ calculated from the pdf $\omega_{d,n}(r)$ (eq. 26)

The moment $\langle R_D^k \rangle$ is written from $\omega_{d,n}(r)$ (eq. 26) as:

$$\langle R_D^k \rangle = \dfrac{\sqrt{\pi}\,\Gamma(nd)}{2^{nd-2}\,\Gamma\left(\dfrac{d}{2}\right)\Gamma\left(\dfrac{n(d+1)}{2}\right)} \times \left\{ \sum_{h=0}^{\lfloor n/2 \rfloor} \dfrac{B(n,h)}{2^h \Gamma(\beta+h)} \int_0^1 r^{d+k-1}\left(1-r^2\right)^{\beta+h-1} dr \right\} \tag{C-8}$$

where $\beta = \dfrac{(n-1)(d-1)}{2}$. The integral in eq. C-8 is readily calculated from the definition of the beta function [24]:

$$\int_0^1 r^{d+k-1}\left(1-r^2\right)^{\beta+h-1} dr = \dfrac{\Gamma\left(\dfrac{d+k}{2}\right)\Gamma(\beta+h)}{2\Gamma\left(\beta+\dfrac{k+d}{2}+h\right)} \tag{C-9}$$

From the Bessel numbers (eq. 23), the bracketed finite sum of eq. C-8 becomes $\dfrac{1}{\Gamma\left(\beta+\dfrac{d+k}{2}\right)}\,{}_2F_1\left(-\dfrac{n}{2},-\dfrac{n}{2}+\dfrac{1}{2};\beta+\dfrac{d+k}{2};1\right)$ because either $-\dfrac{n}{2}$ or $-\dfrac{n}{2}+\dfrac{1}{2}$ is a negative integer for $n \geq 2$. Expressing the latter Gauss hypergeometric function, we obtain finally $(k \geq 0)$:

$$\left\langle R_D^k \right\rangle = \frac{2^k \left(\frac{d}{2}\right)_{k/2} \left(\frac{n(d+1)}{2}\right)_{k/2}}{(nd)_k} \tag{C-10}$$

The moments $\left\langle R_D^k \right\rangle$ are equivalently obtained from $\omega_{d,n}(r) = b_{d,n} r^{d-1} (1-r^2)^{\beta-1} {}_2F_1\left(-\frac{n}{2}, -\frac{n}{2}+\frac{1}{2}; \beta; 1-r^2\right)$ $(r<1)$ (eqs 1 and 27) with $b_{d,n} = \sqrt{\pi} \Gamma(nd) / \left(2^{nd-2} \Gamma(d/2) \Gamma(n(d+1)/2) \Gamma(\beta)\right)$, and from integral 7.512.1 of [24]. The moments, $\left\langle R_D^2 \right\rangle = \frac{d+1}{nd+1}$ and $\left\langle R_D^4 \right\rangle = \frac{(d+1)(d+2)(n(d+1)+2)}{(nd+1)(nd+2)(nd+3)}$ obtained from eq. C-10, are thus proven to be identical with the moments given by eqs C-2 and C-7 respectively.

Eq. C-10 provides an additional check of the validity of the pdf $\omega_{d,n}(r)$ (eq. 26). Indeed, the stochastic representation of the positions of the endpoints of the walks $PG(d,n)$ and $PD(d,n)$, $\mathbf{R}_G \triangleq S\mathbf{R}_D$ (eq. 14) implies that of the distances, $R_G \triangleq SR_D$ and then the equality of the moments:

$$\left\langle R_G^k \right\rangle = \left\langle S^k \right\rangle \left\langle R_D^k \right\rangle \tag{C-11}$$

Both members of eq. C-11 can be calculated independently. Indeed, the gamma distribution $\gamma(nd,1)$ of $S$, the sum of $n$ i.i.d. $\gamma(d,1)$ gamma random variables, implies that $\left\langle S^k \right\rangle = (nd)_k$ (appendix A) so that the right member of eq. C-11 simplifies to $2^k \left(\frac{d}{2}\right)_{k/2} \left(\frac{n(d+1)}{2}\right)_{k/2}$. Further, the moments $\left\langle R_G^k \right\rangle$ are directly calculated from $\psi_{d,n}(r)$ (eq. 6). From integral 6.561.16 [24], these moments are:

$$\begin{cases} v = \dfrac{n(d+1)-d}{2} \\ \left\langle R_G^k \right\rangle = \dfrac{1}{2^{v+d-2}\Gamma\left(\dfrac{d}{2}\right)\Gamma\left(\dfrac{n(d+1)}{2}\right)} \times \int_0^\infty r^{v+d+k-1} K_v(r)\,dr = 2^k \left(\dfrac{d}{2}\right)_{k/2} \left(\dfrac{n(d+1)}{2}\right)_{k/2} \end{cases} \qquad (\text{C-12})$$

The equality of both sides of eq. C-11, which holds if the guessed pdf $p_{d,n}(r)$ is the exact distribution of the position of the endpoint of a $PD(d,n)$ walk, is then established:

$$\left\langle R_G^k \right\rangle = \left\langle S^k \right\rangle \left\langle R_D^k \right\rangle = 2^k \left(\dfrac{d}{2}\right)_{k/2} \left(\dfrac{n(d+1)}{2}\right)_{k/2} \qquad (\text{C-13})$$

The moments of the distribution of the endpoint distance of a $PG(d,n)$ random walk are then such that $\left\langle R_G^k \right\rangle^{-1/(2k)}$ decreases as $1/\sqrt{k}$ when $k$ goes to infinity. The latter distribution, which obeys the Carleman condition for nonnegative random variables, $\sum_{k=1}^{\infty} \left\langle R_G^k \right\rangle^{-1/(2k)} = \infty$, is thus uniquely determined by its sequence of moments which all exist and are finite (Stieltjes moment problem, see for instance [27]).

# Appendix D:
# A relation between modified Bessel functions of the second kind

In proving that the transform eq. 17 connects the guessed density $p_{d,n}(r)$ (eq. 25) to the density $\psi_{d,n}(r)$ (eq. 6) for small values of $n$, we encountered a relation between modified Bessel functions of the second kind $K_\mu(x)$, $(K_\mu(x) = K_{-\mu}(x))$, that we prove below. These functions obey first the relation [20, 24, 28]:

$$K_\mu(x) = \frac{2(\nu-1)}{x} K_{\mu-1}(x) + K_{\mu-2}(x) \tag{D-1}$$

from which it follows that $(\nu > 0,\ m \geq 0)$:

$$K_{\nu+m}(x) = \sum_{k=0}^{m} 2^{m-k} \binom{m}{k} \frac{(\nu)_{m-k}}{x^{m-k}} K_{\nu-k}(x) \tag{D-2}$$

Relation D-2 holds obviously for $m = 0$ and for $m = 1$ (eq. D-1, $\mu = \nu + 1$). To prove it for $m = 2$, it suffices to apply eq. D-1 twice and to write:

$$\begin{cases} K_{\nu+2}(x) = \frac{4\nu(\nu+1)}{x^2} K_\nu(x) + \frac{2(\nu+1)}{x} K_{\nu-1}(x) + \frac{2(\nu-1)}{x} K_{\nu-1}(x) + K_{\nu-2}(x) = \\ \qquad\qquad \frac{4\nu(\nu+1)}{x^2} K_\nu(x) + \frac{4\nu}{x} K_{\nu-1}(x) + K_{\nu-2}(x) \end{cases} \tag{D-3}$$

Assuming that eq. D-2 has been proved true up to $m-1$ inclusive $(m \geq 3)$. $K_{\nu+m}(x)$ writes from equations D-1 and D-2:

$$K_{\nu+m}(x) = \sum_{k=0}^{m-1} 2^{m-k} \binom{m-1}{k} \frac{(\nu+m-1)(\nu)_{m-1-k}}{x^{m-k}} K_{\nu-k}(x)$$

$$+ \sum_{j=0}^{m-2} 2^{m-2-j} \binom{m-2}{j} \frac{(\nu)_{m-2-j}}{x^{m-2-j}} \left\{ \frac{2(\nu-j-1)}{x} K_{\nu-j-1}(x) + K_{\nu-j-2}(x) \right\} \tag{D-4}$$

The multiplying constant $c_{v-j}$ of the function $\dfrac{K_{v-j}(x)}{x^{m-j}}$, $v-2 \leq j \leq v-m+2$, is then:

$$c_{v-j} = 2^{m-j}\left\{\binom{m-1}{j}(v+m-1)(v)_{m-1-j} + \binom{m-2}{j-1}(v-j)(v)_{m-1-j} + \binom{m-2}{j-2}(v)_{m-j}\right\} \quad \text{(D-5)}$$

Consequently :

$$c_{v-j} = \dfrac{2^{m-j}\binom{m}{j}(v)_{m-1-j}}{m(m-1)}\{(v+m-1)(m-j)(m-1)+(v-j)j(m-j)+(v+m-j-1)j(j-1)\} =$$

$$= 2^{m-j}\binom{m}{j}(v)_{m-1-j}(v+m-j-1) = 2^{m-j}\binom{m}{j}(v)_{m-j} \quad \text{(D-6)}$$

The constant $c_v$ which multiplies $\dfrac{K_v(x)}{x^m}$ and that, $c_{v-m}$, which multiplies $K_{v-m}(x)$ are directly found from eq.D-4 to be $2^m(v)_m$ (for $k=0$) and 1 (for $j=m-2$) respectively. Further, the multiplying coefficient $c_{v-1}$ of $\dfrac{K_{v-1}(x)}{x^{m-1}}$ is

$c_{v-1} = 2^{m-1}(v)_{m-2}(v-1+(m-1)(v+m-1)) = 2^{m-1}\binom{m}{1}(v)_{m-1}$. Finally, the coefficient

$c_{v-m+1}$ of $\dfrac{K_{v-m+1}(x)}{x}$ is $c_{v-1} = 2(v+m-1+v-m+1+v(m-2)) = 2mv = 2\binom{m}{m-1}(v)_1$.

Equation D-2 is then seen to hold for $m$ if it holds for $m-1$ and $m-2$. As it is true for $m=1$ and $m=2$ (eqs D-1 and D-3), it is concluded to be valid for any $m$. If $m$ is such that some indices $v-j$ are negative, it suffices to replace the associated Bessel function $K_{v-j}(x)$ by $K_{j-v}(x)$ in eq. D-2. The absolute values of the constant factors in the sum of eq. D-2 are similar to those which appear in a recursive relation established for Bessel functions of the first kind in lemma 2.1 of [18].

# *Appendix E:*
# *The characteristic functions of $R_D$ and $R_G$*

The c.f. $\Phi_{d,n}^{D}(\rho)=\left\langle e^{i\boldsymbol{\rho}.\boldsymbol{R}_D}\right\rangle$ of $\boldsymbol{R}_D$, the endpoint position of a Pearson-Dirichlet random walk, $PD(d,n)$, depends only on the modulus $\rho=\|\boldsymbol{\rho}\|$. As $p_{d,n}(r)$ is a mixture of hyperspherical random walks, the characteristic function $\Phi_{d,n}^{D}(\rho)$ is thus a weighted sum of c.f.'s $\Omega_{k(h)}$ (eq. B-1, appendix B) of uniformly oriented unit vectors of $\mathbb{R}^{k(h)}$. As the latter space dimension is $k(h)=n(d-1)+1+2h$ $\left(h=0,...,\left\lfloor\dfrac{n}{2}\right\rfloor\right)$ from eq. 25, the c.f. is then:

$$\Phi_{d,n}^{D}(\rho)=\sum_{h=0}^{\left\lfloor\frac{n}{2}\right\rfloor} w_{d,n}(h)\,\Omega_{n(d-1)+1+2h}(\rho) \qquad (\text{E-1})$$

From the weights $w_{d,n}(h)$ given by eq. 24 and from eq. B-2, the previous c.f. is explicitly written as:

$$\begin{cases} \Phi_{d,n}^{D}(\rho)=Z_{d,n}\left\{\sum_{h=0}^{\left\lfloor\frac{n}{2}\right\rfloor} B(n,h)\times\dfrac{J_{\frac{n(d-1)-1}{2}+h}(\rho)}{\rho^{\frac{n(d-1)-1}{2}+h}}\right\} \\[2ex] Z_{d,n}=2^{\frac{n(d-1)-1}{2}}\times\dfrac{\Gamma\!\left(\dfrac{nd}{2}\right)\Gamma\!\left(\dfrac{nd+1}{2}\right)}{\Gamma\!\left(\dfrac{n(d+1)}{2}\right)} \end{cases} \qquad (\text{E-2})$$

The endpoint $\boldsymbol{R}_G$ of a Pearson-Gamma random walk, $PG(d,n)$, is related to $\boldsymbol{R}_D$ by the stochastic equation, $\boldsymbol{R}_G \triangleq S\boldsymbol{R}_D$ (eq. 14). Then, the characteristic function $\Phi_{d,n}^{G}(\rho)$ of $\boldsymbol{R}_G$ is calculated from the gamma distribution $\gamma(nd,1)$ of the total length $S$ and from the c.f. $\Phi_{d,n}^{D}(\rho)$ to be:

$$\Phi_{d,n}^{G}(\rho)=\dfrac{1}{\Gamma(nd)}\int_{0}^{\infty} s^{nd-1}e^{-s}\Phi_{d,n}^{D}(\rho s)\,ds = \dfrac{1}{\rho^{nd}\Gamma(nd)}\int_{0}^{\infty} x^{nd-1}e^{-x/\rho}\Phi_{d,n}^{D}(x)\,dx \qquad (\text{E-3})$$

The c.f. $\Phi_{d,n}^G(\rho)$ is readily found to be equal to $\dfrac{1}{\left(1+\rho^2\right)^{n(d+1)/2}}$ (eq. 5). We aim to show below that the latter c.f. is indeed retrieved when eq. E-2 is plugged into eq. E-3:

$$\Phi_{d,n}^G(x) = Z_{d,n}\left\{\sum_{h=0}^{\lfloor\frac{n}{2}\rfloor} B(n,h)\int_0^\infty x^{\frac{n(d-1)-1}{2}+h}\exp\left(-\frac{x}{\rho}\right) J_{\frac{n(d-1)-1}{2}+h}(x)\,dx\right\} \qquad (E\text{-}4)$$

From integral 6.621.1 of [24], the integral in eq. E-4 writes:

$$\int_0^\infty x^{\frac{n(d-1)-1}{2}+h}\exp\left(-\frac{x}{\rho}\right) J_{\frac{n(d-1)-1}{2}+h}(x)\,dx =$$

$$= \frac{\Gamma(nd)}{2^{\frac{n(d-1)-1}{2}+h}\Gamma\left(\frac{n(d-1)+1}{2}+h\right)} \times \rho^{nd}\,{}_2F_1\left(\frac{nd}{2},\frac{n(d+1)}{2};\frac{n(d-1)+1}{2}+h\,;\,-\rho^2\right) \qquad (E\text{-}5)$$

where we have further used the relation ${}_2F_1(\alpha,\beta;\gamma;z) = (1-z)^{-\alpha}\,{}_2F_1\left(\alpha,\gamma-\beta;\gamma;\dfrac{z}{z-1}\right)$

(eq. 9.131.1 of [24]). Then:

$$\begin{cases}\Phi_{d,n}^G(\rho) = T_{d,n}\left\{\displaystyle\sum_{h=0}^{\lfloor\frac{n}{2}\rfloor}\frac{B(n,h)}{2^h\Gamma\left(\frac{n(d-1)+1}{2}+h\right)}\times {}_2F_1\left(\frac{nd}{2},\frac{nd+1}{2};\frac{n(d-1)+1}{2}+h\,;-\rho^2\right)\right\}\\[2ex]T_{d,n} = \dfrac{\Gamma\left(\frac{nd}{2}\right)\Gamma\left(\frac{nd+1}{2}\right)}{\Gamma\left(\frac{n(d+1)}{2}\right)}\end{cases} \qquad (E\text{-}6)$$

Replacing the Gauss hypergeometric function in eq. E-6 by an infinite sum [24]:

$${}_2F_1\left(\frac{nd}{2},\frac{nd+1}{2};\frac{n(d-1)+1}{2}+h\,;-\rho^2\right) = \sum_{k=0}^\infty \frac{\left(\frac{nd}{2}\right)_k \left(\frac{nd+1}{2}\right)_k}{\left(\frac{n(d-1)+1}{2}+h\right)_k}\times\frac{(-1)^k\rho^{2k}}{k!} \qquad (E\text{-}7)$$

then expressing the Bessel numbers $B(n,h)$ as $\dfrac{2^h}{h!} \times \left(-\dfrac{n}{2}\right)_h \left(-\dfrac{n}{2}+\dfrac{1}{2}\right)_h$ (eq. 23) and finally collecting, for every $k$, the $\left\lfloor \dfrac{n}{2} \right\rfloor$ terms in $\rho^{2k}$, we rewrite eq. E-6 as:

$$\Phi_{d,n}^{G}(\rho) = T_{d,n} \sum_{k=0}^{\infty} \frac{\left(\dfrac{nd}{2}\right)_k \left(\dfrac{nd+1}{2}\right)_k}{\Gamma\left(\dfrac{n(d-1)+1}{2}+k\right)} \times \frac{(-1)^k \rho^{2k}}{k!} \times \left[ \sum_{h=0}^{\left\lfloor \frac{n}{2} \right\rfloor} \frac{\left(-\dfrac{n}{2}\right)_h \left(-\dfrac{n}{2}+\dfrac{1}{2}\right)_h}{\left(\dfrac{n(d-1)+1}{2}+k\right)_h h!} \right] \quad \text{(E-8)}$$

The bracketed sum in eq. E-8 is a Gauss hypergeometric function whose series terminates in all cases (section 5.2). Its value is then:

$$_2F_1\left(-\dfrac{n}{2}, -\dfrac{n}{2}+\dfrac{1}{2}; \dfrac{n(d-1)+1}{2}+k; 1\right) = \frac{\Gamma\left(\dfrac{n(d-1)+1}{2}+k\right)\Gamma\left(\dfrac{n(d+1)}{2}+k\right)}{\Gamma\left(\dfrac{nd}{2}+k\right)\Gamma\left(\dfrac{nd+1}{2}+k\right)} \quad \text{(E-9)}$$

As $_2F_1(\alpha, \beta; \beta; z) = (1-z)^{-\alpha}$ for arbitrary $\alpha$ (eq. 15.4.6 of [20]), the c.f. $\Phi_{d,n}^{G}(\rho)$ given by eq. E-8 simplifies finally to:

$$\Phi_{d,n}^{G}(\rho) = \sum_{k=0}^{\infty} \left(\dfrac{n(d+1)}{2}\right)_k \times \frac{(-\rho^2)^k}{k!} = \frac{1}{(1+\rho^2)^{n(d+1)/2}} \quad \text{(E-10)}$$

In conclusion, the characteristic function $\Phi_{d,n}^{G}(\rho)$ calculated from eqs E-2 and E-3 is that calculated directly for Pearson-Gamma random walks (eq. 5).

As eq. E-3 is a Laplace transform, $L(f(t)) = \int_0^{\infty} f(t) e^{-st} dt$, the previous calculation yields in addition the following inverse Laplace transform $(s > 0)$:

$$L^{-1}\left(\frac{s^n}{(1+s^2)^{nm/2}}\right) = \frac{\sqrt{\pi}}{2^{\frac{nm-1}{2}} \Gamma\left(\dfrac{nm}{2}\right)} \left\{ \sum_{h=0}^{\left\lfloor \frac{n}{2} \right\rfloor} B(n,h) t^{\frac{nm-1}{2}-h} J_{\frac{n(m-2)-1}{2}+h}(t) \right\} \quad \text{(E-11)}$$

which holds at least for integer values of $n$ and $m$ larger respectively than 1 and 2.

**_Table 1:_** Weights $w_{d,n}(h)$, $h = 0,..,\left\lfloor \dfrac{n}{2} \right\rfloor$, of the densities which sum up to give the total density $p_{d,n}(r)$ (eq. 19) of the endpoint position of a $PD(d,n)$ random walk of $2 \leq n \leq 5$ steps in $\mathbb{R}^d$.

| $n$ | $w_{d,n}(0)$ | $w_{d,n}(1)$ | $w_{d,n}(2)$ |
|---|---|---|---|
| 2 | $\dfrac{2d-1}{2d}$ | $\dfrac{1}{2d}$ | - |
| 3 | $\dfrac{3d-2}{3d+1}$ | $\dfrac{3}{3d+1}$ | - |
| 4 | $\dfrac{(4d-3)(4d-1)}{4d(4d+2)}$ | $\dfrac{6(4d-1)}{4d(4d+2)}$ | $\dfrac{3}{4d(4d+2)}$ |
| 5 | $\dfrac{(5d-4)(5d-2)}{(5d+1)(5d+3)}$ | $\dfrac{10(5d-2)}{(5d+1)(5d+3)}$ | $\dfrac{15}{(5d+1)(5d+3)}$ |

**_Table 2:_** The $d$-independent factors, $B(n,h)$ of the weights

$$w_{d,n}(h) = B(n,h) c_{d,n}(h), \quad h = 0,..,\left\lfloor \dfrac{n}{2} \right\rfloor \text{ for } 4 \leq n \leq 9.$$

| $n$ | $B(n,0)$ | $B(n,1)$ | $B(n,2)$ | $B(n,3)$ | $B(n,4)$ |
|---|---|---|---|---|---|
| 4 | 1 | 6 | 3 | - | - |
| 5 | 1 | 10 | 15 | - | - |
| 6 | 1 | 15 | 45 | 15 | - |
| 7 | 1 | 21 | 105 | 105 | - |
| 8 | 1 | 28 | 210 | 420 | 105 |
| 9 | 1 | 36 | 378 | 1260 | 945 |

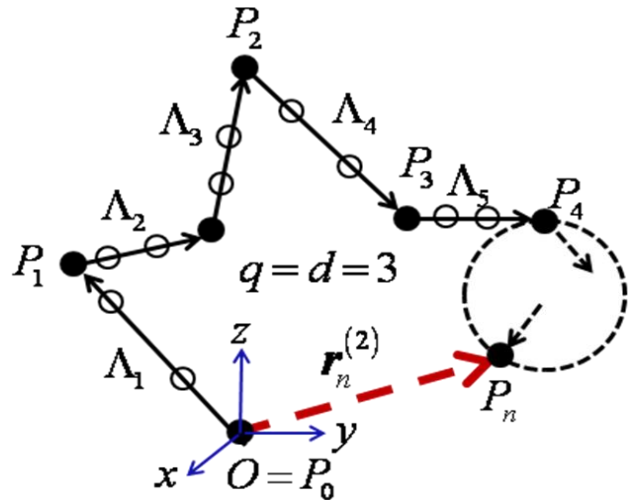

***Figure 1:*** A Pearson-Gamma random walk $PG(3,n)$ in $\mathbb{R}^3$ for $q=d=3$: the walk starts at O in a random direction; the length of every step $\Lambda_k = S_k \, (k=1,..,n)$ is gamma distributed, $\gamma(d,1)$, being the sum of $d$ (as indicated by $d-1$ empty circles) i.i.d. exponential random variables; at $P_k \, (k=1,..,n-1)$, a new random direction is taken independently of the previous ones. For the associated Pearson-Dirichlet random walk $PD(3,n)$, every step $\Lambda_k = L_k$ is rescaled so as to make the total travelled length equal to 1 $\left( \sum_{k=1}^{n} \Lambda_k = 1 \right)$. The walk ends at $P_n$.

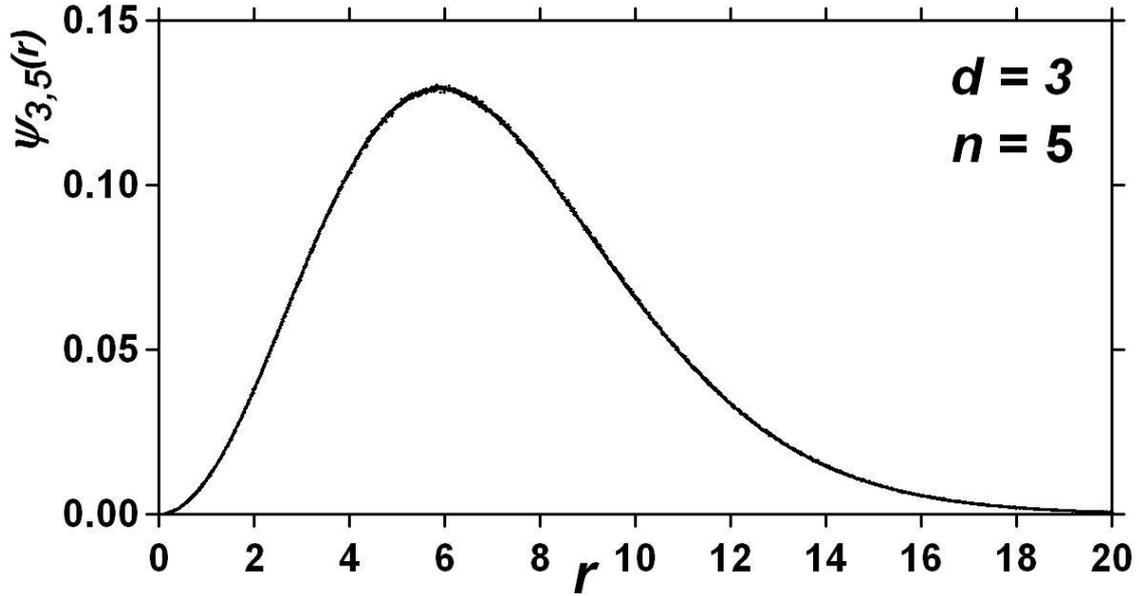

*__Figure 2:__* Monte-Carlo simulation ($10^8$ walks) of a Pearson-Gamma random walk of 5 steps in $\mathbb{R}^3$ with $q = d = 3$. The pdf of each step length is $p(s_k) = s_k^2 e^{-s_k}/2 \ (s_k > 0, \ k = 1,..,5)$. The simulated pdf $\psi_{3,5}(r)$ (points) is compared to the calculated one: $\psi_{3,5}(r) = \dfrac{r^{21/2} K_{17/2}(r)}{92897280\sqrt{2\pi}}$ (solid line, eq. 6).

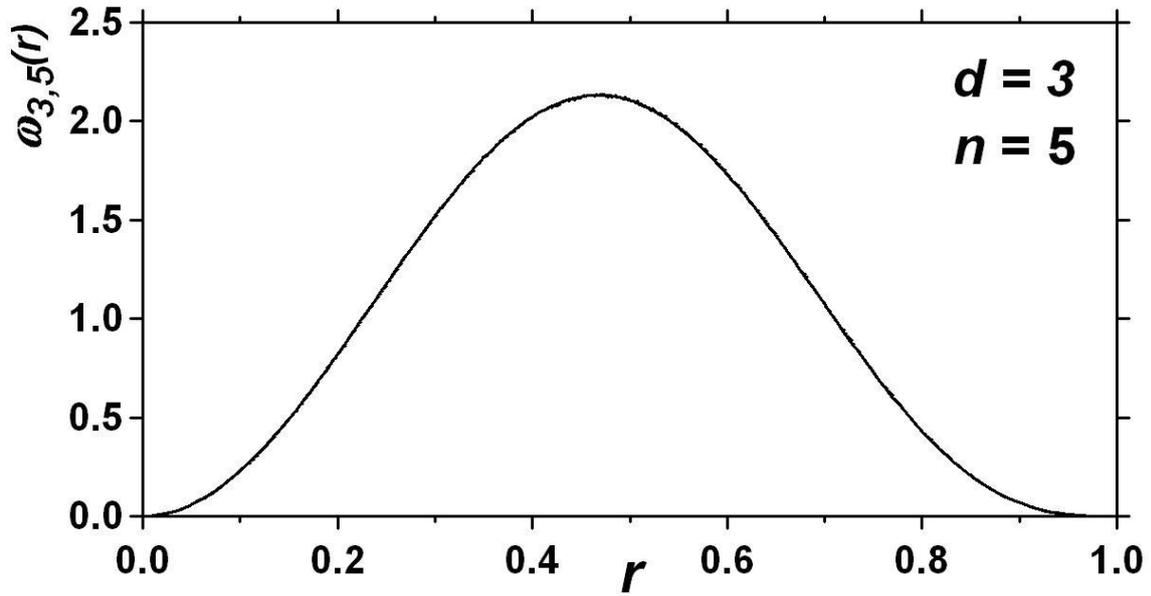

***Figure 3:*** Monte-Carlo simulation ($10^8$ walks) of a Pearson-Dirichlet random walk of 5 steps in $\mathbb{R}^3$ with $q = d = 3$. The simulated pdf $\omega_{3,5}(r)$ (points) is compared to the calculated one: $\omega_{3,5}(r) = \dfrac{5005}{8192} r^2 (1-r^2)^3 (39 - 26r^2 + 3r^4)$ (solid line, eq.26).

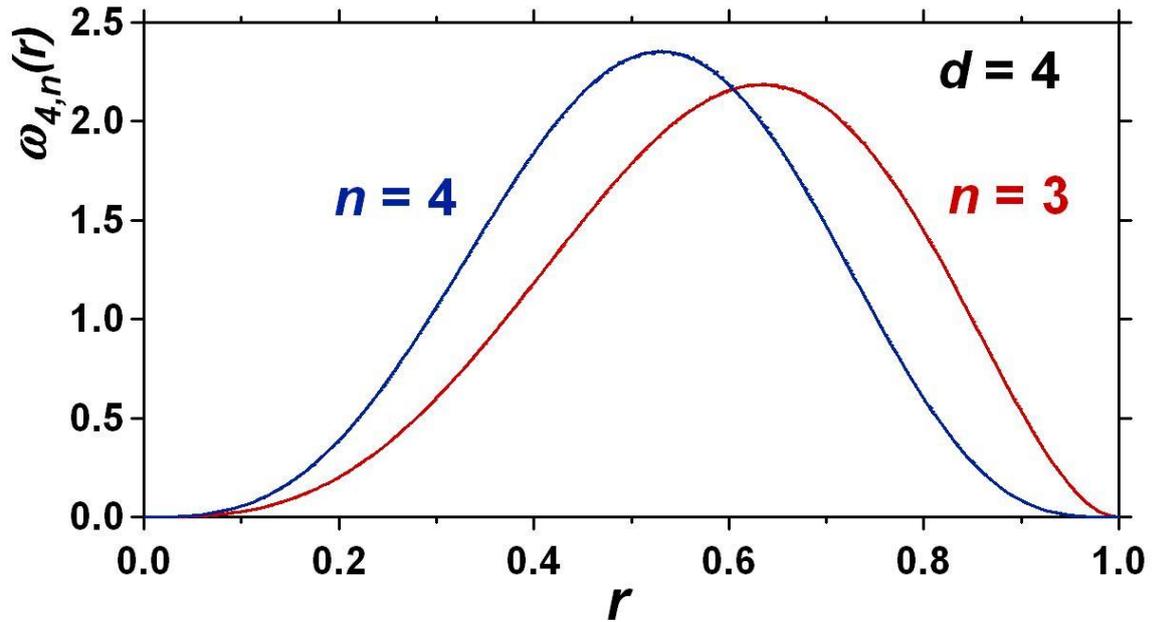

***Figure 4:*** (color on-line) Monte-Carlo simulations ($10^8$ walks) of Pearson-Dirichlet random walks of $n = 3, 4$ steps in $\mathbb{R}^4$ with $q = d = 4$. The simulated pdf's $\omega_{4,n}(r)$ (points) are compared to the calculated ones: $\omega_{4,3}(r) = \dfrac{120}{13} r^3 (1-r^2)^2 (3-r^2)$ and $\omega_{4,4}(r) = \dfrac{65}{64} r^3 (1-r^2)^{7/2} (56 - 24r^2 + r^4)$ (solid lines, eq. 26).